\documentclass[runningheads]{llncs}
\usepackage[T1]{fontenc}
\usepackage{graphicx}
\usepackage{amsmath, amsfonts, amssymb}
\usepackage{bm}
\usepackage{subfigure}
\usepackage{multirow}
\begin{document}
\title{Identifying Suspicious Regions of COVID-19 by Abnormality-Sensitive Activation Mapping}
\titlerunning{Suspicious-region identification}
%
\author{
  Ryo Toda\inst{1}\and
  Hayato Itoh\inst{1} \and
  Masahiro Oda\inst{2,1} \and
  Yuichiro Hayashi\inst{1} \and
  Yoshito Otake\inst{3,4} \and
  Masahiro Hashimoto\inst{5} \and
  Toshiaki Akashi\inst{6} \and
  Shigeki Aoki\inst{6} \and
  Kensaku Mori\inst{1,4,7}
}

\authorrunning{R. Toda et al.}

%
\institute{
  Graduate School of Informatics, Nagoya University, Nagoya, Japan\and
  Information and Communications, Nagoya University, Nagoya, Japan \and
  Graduate School of Science and Technology, Nara Institute of Science and Technology,
  Nara, Japan \and
  Research Center for Medical BigData, National Institute of Informatics, Tokyo, Japan \and
  Department of Radiology, Keio University School of Medicine, Tokyo, Japan \and
  Department of Radiology, Juntendo University, Tokyo, Japan \and
  Information Technology Center, Nagoya University, Nagoya, Japan
}
\maketitle              
\begin{abstract}
This paper presents a fully-automated method for the identification of suspicious regions of
a coronavirus disease (COVID-19) on chest CT volumes.
One major role of chest CT scanning in COVID-19 diagnoses is
identification of an inflammation particular to the disease.
This task is generally performed by radiologists through an interpretation of the CT volumes,
however, because of the heavy workload,
an automatic analysis method using a computer is desired.
Most computer-aided diagnosis studies have addressed only a portion of the
elements necessary for the identification.
In this work, we realize the identification method through a classification task 
by using a 2.5-dimensional CNN with three-dimensional attention mechanisms.
We visualize the suspicious regions by applying a backpropagation based on positive gradients to
attention-weighted features.
We perform experiments on an in-house dataset and two public datasets
to reveal the generalization ability of the proposed method.
The proposed architecture achieved AUCs of over 0.900 for all the datasets, and 
mean sensitivity $0.853 \pm 0.036$ and specificity $0.870 \pm 0.040$.  
The method can also identify notable lesions pointed out in the radiology report
as suspicious regions.
\end{abstract}

\section{Introduction}
The identification of suspicious abnormality from a medical image is
a fundamental problem with relevance to many computer-aided diagnosis (CAD) systems.
Automatic and precise identification of suspicious regions remains a major challenge in the medical image processing field. 
Facing at an infectious disease caused by the coronavirus SARS-CoV-2 (COVID-19), 
chest CT scanning is an essential examination for COVID-19 diagnosis \cite{CTonCOVID,CTonCOVID2,CTusefulness}, and 
automatic identification of suspicious region from a CT volume also poses high
demands on image analysis to reduce interpreting workloads for radiologists \cite{COVIDAIUseless}.
Along with increasing demands for early COVID-19 diagnoses,
numerous automated classification methods have been investigated \cite{COVIDClassification,EarlyCOVIDAI2,EarlyCOVIDAI3,EMARS}.
However, these works focused on the precise classification of COVID-19 cases \cite{COVIDCNN,COVIDCNN2}.
Toward practical application to medical diagnosis, in addition to high
classification accuracy, explainability/interpretability by a suggestion
of suspicious regions found by classification model is essential \cite{COVIDAIUseless}. 

Related works addressed automated identification of suspicious regions in COVID-19 diagnosis.
Han et al. \cite{AttentionMIL} proposed 3D attention-based multiple instance learning and they visualized 3D patch-level importance.
This method suggests only binary information whether suspicious regions or not for small 3D patches in a CT volume.
Zhang et al. applied attention blocks for fusing 3D chest CT volume and 2D
X-ray image and visualized suspicious regions \cite{AttentionGrad}
Takateyama et al. \cite{TUAT} utilized attention specific to bilateral
lesions and pleural effusions and visualized saliency in input. 
These two methods simply apply 2D Grad-CAM to their trained models.
Because these slice-level visualizations typically show high intensities
for normal structures that span multiple slices such as bronchi, blood
vessels, and interlobular septa, these are insufficient for identifying key patterns. 
Therefore, visualizing 3D suspicious lesions is still an open problem. 
Our motivation is to reveal COVID-19 suspicious regions by utilizing 3D information effectively.

Visual explanation of key patterns realizes the identification of the suspicious regions of COVID-19.
Since this approach bases on the assumption that a trained model can
correctly recognize key patterns and accurately classify chest CT volumes, 
we propose 2.5D architecture for robust COVID-19 classification.
This hybrid architecture of three 2D encoders followed by a single 3D encoder extracts fine 3D features 
with a reduced number of parameters. 
Furthermore, we incorporate attention mechanisms into our 2.5D architecture to capture key 3D patterns. 
While low-resolution attention maps given by attention mechanisms
roughly imply regions including the findings of target COVID-19 disease,
it is difficult to understand where and what are actual key patterns at the original-scale images/volumes.
To circumvent this difficulty, we extend a backpropagation-based visual
explanation method for a 3D convolutional layer and applied it to our attention-guided 3D features towards the identification of suspicious regions.

In this work, we propose a method for the identification of COVID-19 suspicious
regions by introducing abnormality-sensitive activation mapping. 
For the visualization of suspicious regions,
this method selectively utilize features relevant for the classification.
A summary of our main contributions are follows:
(1) we offer an attention-guided full-3D visual explanation method toward the COVID-19 diagnostic assistance,
(2) our identification provides a strong interpretability for the classification results by directly pointing out actual lesions, and
(3) we perform large-scale experiments on three datasets, with a total of over 2,000 cases.

\section{Methods}
\begin{figure}[tb]
  \centering
  \includegraphics[width=\textwidth]{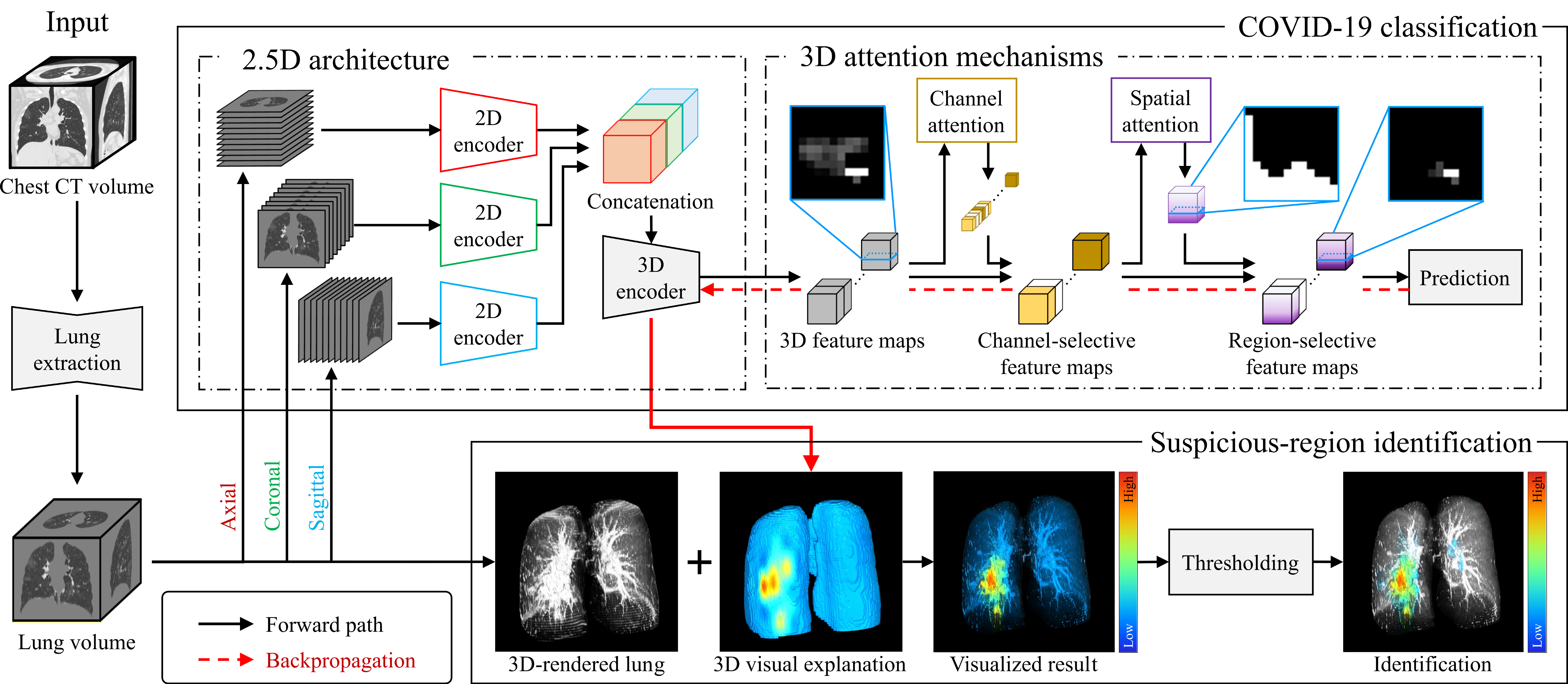}
  \caption{
    Outline of the proposed method.
    We create a robust model for automated classification of COVID-19 and
    identify suspicious regions of COVID-19 by analyzing the model.
  }\label{fig:Outline}
\end{figure}

Our proposed method consists of a robust 2.5D classifier with our extended attention mechanisms and its visual explanation.
Since we integrated the feature extraction based on the extended 3D attention mechanisms,
we can assume that larger gradients backpropagate to only key features
in each layer.
Therefore, backpropagation via attention-guided features leads our abnormality-sensitive activation mapping. 
The input and output of the proposed method is a chest CT volume
$\bm{V}_{\mathrm{CT}}\in\mathbb{R}^{512\times512\times D}$ and
a 3D visual explanation showing suspicious regions, respectively.
Figure \ref{fig:Outline} shows the outline of our method.

\subsection{Robust COVID-19 classification}
\subsubsection{Preprocessing}
We set a window level and a window width of $\bm{V}_{\mathrm{CT}}$ to $-$550 H.U. and 1500 H.U.,
respectively, for a lung window setting.
We normalize the volume $\bm{V}_{\mathrm{CT}}$ into
$\bm{\hat{V}}_{\mathrm{CT}}\in{[-1,1]}^{512\times512\times D}$.
After the normalization, a lung segmentation method \cite{SegModel} processes the volume
to output a lung mask volume $\bm{V}_{\mathrm{m}}\in\{0,1\}^{512\times512\times D}$,
where 0 and 1 express outside and inside the lung, respectively.
By using an element-wise multiplication of $\bm{\hat{V}}_{\mathrm{CT}}$ and $\bm{V}_{\mathrm{m}}$,
we obtain a lung volume $\bm{V} = \bm{\hat{V}}_{\mathrm{CT}} \odot \bm{V}_{\mathrm{m}}$.

\subsubsection{2.5D-based representation learning}
We design a 2.5D architecture for effective 3D feature extraction from 3D volumetric CT data.
Since 3D volumes can express more patterns than 2D images,
an architecture of 3D CNN also have more and more parameters than one of 2D CNN, and
apt to result in overfitting to a given training dataset \cite{3Doverfit1,3Doverfit2}. 
However, our hybrid architecture of three 2D encoders and one 3D encoder
can avoid overfitting to a training dataset by reducing the number of parameters of CNN.

For a resized lung volume $\bm{\hat{V}}\in[-1,1]^{192\times192\times64}$ of $\bm{V}$,  
three 2D encoders of four convolutional layers perform 2D convolutions with $3\times3$ kernels 
along axial, coronal, and sagittal slices of $\bm{\hat{V}}$, respectively.
As a result of these convolutions, we obtain tensors
$\bm{F}_{\mathrm{ax}}, \bm{F}_{\mathrm{cor}}, \bm{F}_{\mathrm{sag}}
\in\mathbb{R}^{32\times48\times48\times16}$.
By concatenating these three feature maps, we obtain a tensor 
$\bm{F}_{\mathrm{con}}\in\mathbb{R}^{96\times48\times48\times16}$.
For $\bm{F}_{\mathrm{con}}$, the 3D encoder performs 3D convolutions
with $3\times3\times3$ and $1\times1\times1$ kernels accompanying
dilated convolutions \cite{Dilated} and mixed pooing \cite{MixedPool}.
Concatenated 2D features are fused into 3D feature maps
$\bm{F}_c \in \mathbb{R}^{12 \times 12 \times 4}$ with a channel index $c=1,2, \dots, 256$.

\subsubsection{Attention-guided 3D feature selection}
We develop attention mechanisms for 3D feature maps by extending channel and spatial attentions \cite{CBAM}.
These mechanisms enable us to detect regions including key patterns for accurate classification. 
For 3D feature maps $\bm{F}_c \in \mathbb{R}^{W \times H \times D}$ with
a channel index $c=1,2, \dots, C$,  we compute spatial-wise
global-average and max poolings by  
\begin{equation}
  a_c=\frac{1}{WHD}\sum_{x,y,z=1}^{W,H,D}F_{cxyz}, \, \, \, \, 
   b_c=\max
   (F_{c111},F_{c112}, \ldots, F_{cWHD}),
 \end{equation}
respectively, and then obtain vectors $\bm{a}=(a_c), \bm{b} =(b_c) \in \mathbb{R}^C$.
By feeding these $\bm{a}$ and $\bm{b}$ into two multi-layer perceptrons 
of weights $\bm{W}_0$ and $\bm{W}_1$ with different activation functions, 
we compute a $C$-dimensional attention vector $\bm{m} = (m_c) \in[0,1]^C$ as 
\begin{equation}
  \label{eq:ChannelAtt}
  \begin{split}
    \bm{m} = \bm{\sigma}(\bm{W}_1^{\top}\bm{\rho}(\bm{W}_0^{\top}\bm{a})
    + \bm{W}_1^{\top}\bm{\rho}(\bm{W}_0^{\top}\bm{b})),
  \end{split}
\end{equation}
where $\bm{\sigma}$ and $\bm{\rho}$ express element-wise activations by sigmoid and ReLU functions, respectively.
From $\bm{m}$ and $\{\bm{F}_c\}_{c=1}^C$, we obtain intermediate feature
maps $\bm{F}'_c = m_c\bm{F}_c$ for $c=1, 2, \dots, C$.

For the computation of spatial attention, channel-wise average- and
max-poolings of $\bm{F}'_c = ({F}'_{cxyz})$ generate tensors
$\bm{A}=(A_{xyz}), \bm{B}=(B_{xyz})\in\mathbb{R}^{W \times H \times D}$, respectively, by
\begin{equation}
A_{xyz}=\frac{1}{C}\sum_{c=1}^CF'_{cxyz}, \, \, \, 
B_{xyz}=\max(F'_{1xyz},F'_{2xyz},\ldots,F'_{Cxyz}).
\end{equation}
Concatenating these two tensors as
$\bm{C}=[\bm{A},\bm{B}]\in\mathbb{R}^{W \times H \times D\times 2}$ and
setting a 3D convolution as
$\bm{C}*\bm{K}=(\bm{\hat{C}}_{xyz}),$ where $\bm{K}\in\mathbb{R}^{3\times3\times3}$
is a convolution kernel,
we have a spatial-attention map $\bm{M}=(M_{xyz}) \in[0,1]^{W\times
H\times D}$ by
\begin{equation}
  \begin{split}
  M_{xyz}=\sigma(\hat{C}_{xyz}),
  \end{split}
\end{equation}
where $\sigma$ expresses a sigmoid activation function.
Using element-wise multiplication of two tensors,
we obtain attention-guided feature maps by
\begin{equation}
  \label{eq:selectedF}
  \bm{F}''_c=\bm{M}\odot\bm{F}'_c,
\end{equation}
as the final output of our two attention mechanisms.
By using these feature maps, our model computes likelihoods $y^{l}$ for two
classes and then outputs a predicted class $l \in \{ 0,1 \}$,
where 0 and 1 indicate non-typical and typical cases, respectively.

\subsection{Identification of suspicious region via weight analysis}
Visual explanation of our robust COVID-19 classification model realizes the identification of suspicious regions. 
To achieve the identification of suspicious regions, 
we propose a positive-gradient-based visual explaining method for 3D patterns.
While the survey \cite{XAIinMedicine} of deep learning methods in medical imaging analysis
reported that many studies used the Grad-CAM \cite{GradCAM}, 
recent works indicate the importance of positive gradients
in the weight analysis of a trained CNN \cite{PosiGradCAM,PosiEqualPP}.
Furthermore, mathematical analysis in Ref. \cite{PosiEqualPP} reveals
that the numerical equivalence of Grad-CAM++ \cite{GradCAMpp}
and positive-gradient-based Grad-CAM \cite{PosiGradCAM}.

We set $y^{(l)}$ as an output of a CNN for class $l$.
For each voxel $v_{cxyz}$ in $c$-th 3D feature map of size $I \times J \times K$
at a 3D convolutional layer, 
we have a neuron importance weight
\begin{equation}
    \label{eq:Weight3D}
    \alpha_c^{(l)} = \frac{1}{IJK}\sum_{x=1}^I \sum_{y=1}^J \sum_{z=1}^K
    \mathrm{ReLU}\left( \frac{\partial y^{(l)}}{\partial v_{cxyz}} \right),
\end{equation}
where $\partial y^{(l)}/\partial v_{cxyz}$ is
a gradient of $y^{(l)}$ with respect to $v_{cxyz}$.
Visual explanation $\bm{S}^{(l)} = ( s_{xyz}^{(l)} ) \in \mathbb{R}^{I
\times J \times K}$  at a 3D convolutional layer for class $c$ is given
by
\begin{equation}
    \label{eq:PosiGradCAM3D}
    s_{xyz}^{(l)} = \mathrm{ReLU}
    \left(
      \sum_{c=1}^C \alpha_c^{(l)} v_{cxyz}
    \right).
\end{equation}
We apply Eqs. (\ref{eq:Weight3D}) and (\ref{eq:PosiGradCAM3D})
to the convolutional layer just before the final down sampling.
In order to clarify the suspicious regions, we exclude parts of the heatmap, which have intensities of 0.1 or less, from the
final identification result.

\section{Experiments}
\subsubsection{Datasets}
\begin{table}[b]
  \caption{
    Details of three datasets.
  }
  \label{tab:Dataset}
  \centering
  \begin{tabular}{c|c||r|r|r}
    \hline
    \multicolumn{2}{c||}{Dataset} & \multicolumn{1}{|c|}{Typical} &
    \multicolumn{1}{c|}{Non-typical} & \multicolumn{1}{c}{Total}\\
    \hline
    \multirow{4}{*}{In-house} & Training & 596 & 368 & 964\\
    & Validation & 149 & 92 & 241\\
    & Testing & 187 & 115 & 302\\
    & All & 932 & 575 & 1,507\\
    \hline
    COVID-CT-MD \cite{Dataset1} & Testing & 169 & 76 & 245 \\
    COVID-CTset \cite{Dataset2} & Testing & 95 & 282 & 377 \\
    \hline
  \end{tabular}
\end{table}
To evaluate the proposed method, we used an IRB-approved in-house dataset and
two public datasets \cite{Dataset1,Dataset2}.
Table \ref{tab:Dataset} summarizes the details of datasets.
For the in-house dataset, radiologists labeled each case based on
the four COVID-19 typicality categorization defined by the RSNA \cite{RSNA}.
From the given labels, we used two high typicality categories as the
typical COVID-19 and the rest as the non-typical case.
Even though these public datasets offer labels of COVID-19 and normal cases,
we treated the former as typical and the latter as non-typical.

\subsubsection{Implimentation and training}
Our implementation is based on TensorFlow2.
For computation, we used two AMD EPYC7313 processors and an NVIDIA A100 GPU of 80 GB RAM.
We trained the 2.5D CNN with 3D attention mechanisms for 100 epochs using Adam optimizer
with an initial learning rate of $1.0\times10^{-4}$.
The learning rate decayed 15\% at every 10 epochs.  The batch size was 16.
We selected the best models of each method w.r.t. classification accuracy in the validation dataset.

\subsubsection{Evaluation of COVID-19 classification}
\begin{figure}[tb]
    \centering
    \subfigure[]{
      \includegraphics[width=0.40\textwidth]{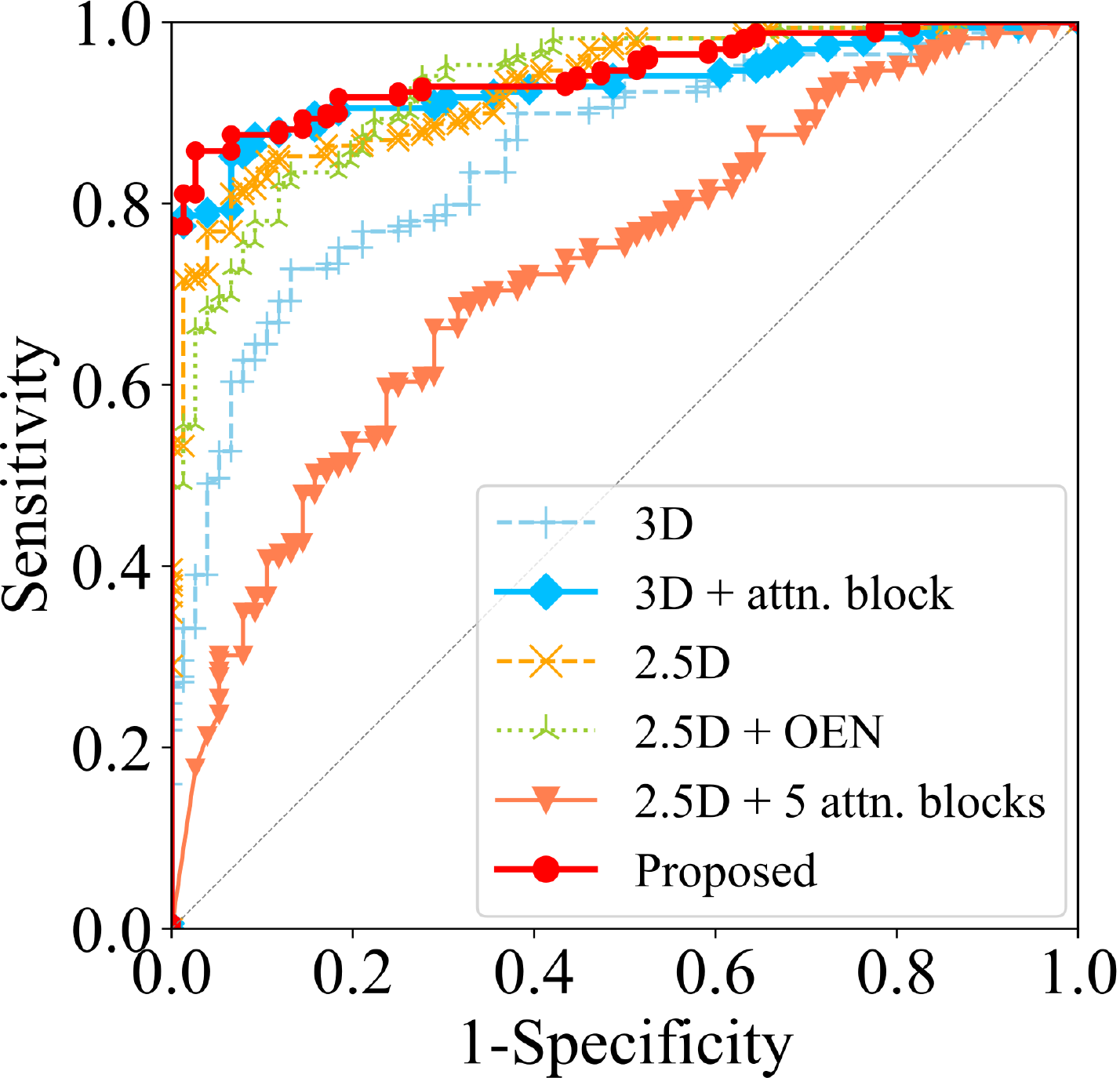}
    }
    \subfigure[]{
      \includegraphics[width=0.40\textwidth]{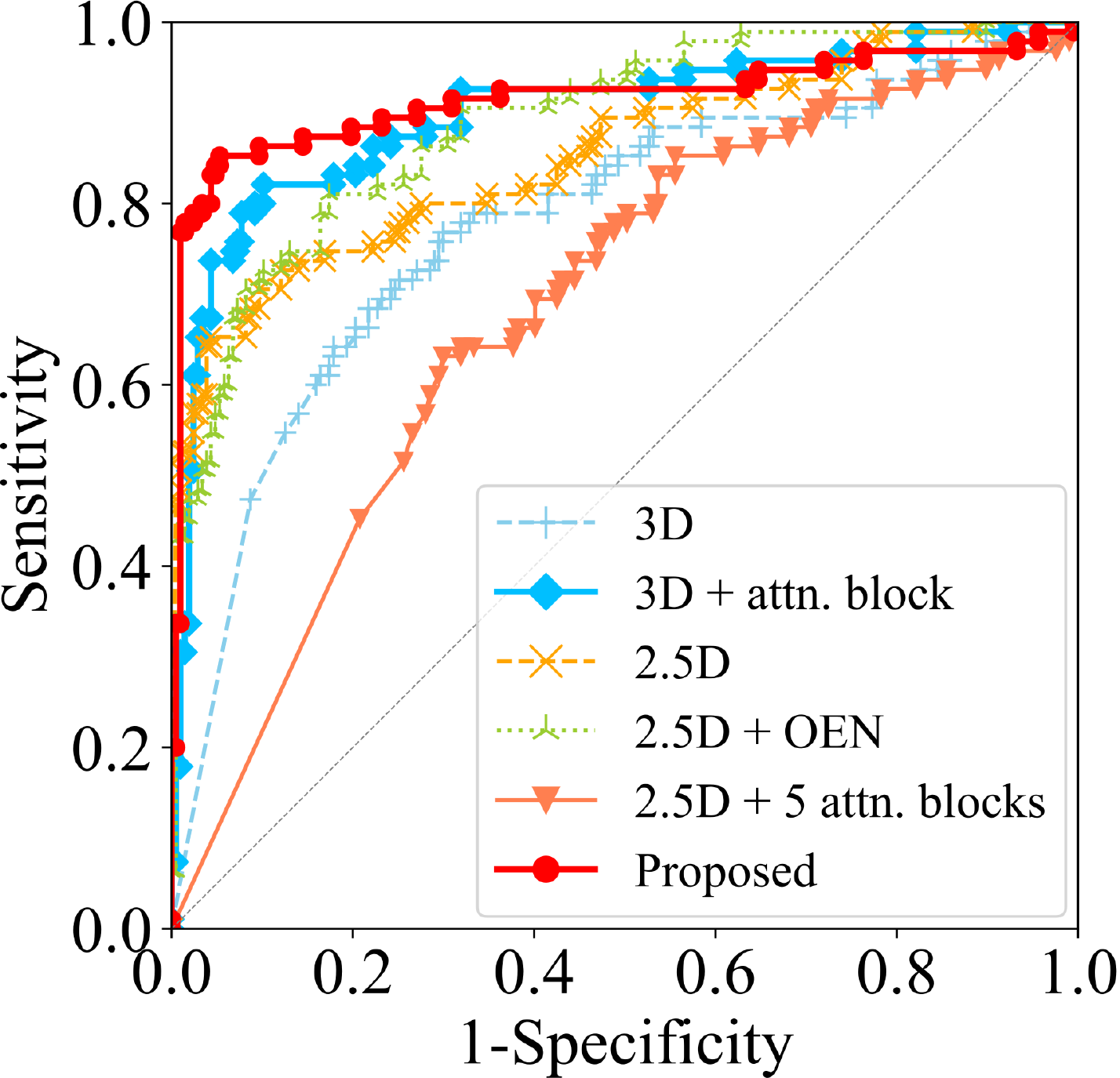}
    }
    \caption{
        ROC curves of six methods for two public datasets:
        (a) COVID-CT-MD and (b) COVID-CTset.
        These plots are based on a sensitivity and a specificity obtained
        by varying a threshold that bound typical and non-typical cases
        for the output likelihoods of each method.}\label{fig:ROCCurves}
\end{figure}

We evaluated the 2.5D CNN's performance by using
receiver operating characteristic (ROC) curves and areas under the curves (AUCs)
based on the model's classification accuracy.
We compared the proposed model with a 3D CNN \cite{CNNModel},
a 3D CNN with one attention block,
a 2.5D CNN w/o attention, an orthogonal ensemble \cite{OEN} of 2.5D CNNs
and a 2.5D CNN with five attention blocks. 
Figure \ref{fig:ROCCurves} and Table \ref{tab:AUCs} show ROC curves and AUCs
based on the classification results for each dataset.
The proposed architecture achieved the best AUCs on three datasets, and 
mean sensitivity $0.853 \pm 0.036$ and specificity $0.870 \pm 0.040$.

\subsubsection{Evaluation of suspicious region identification}
\begin{figure}[tb]
    \centering
    \includegraphics[width=0.9\textwidth]{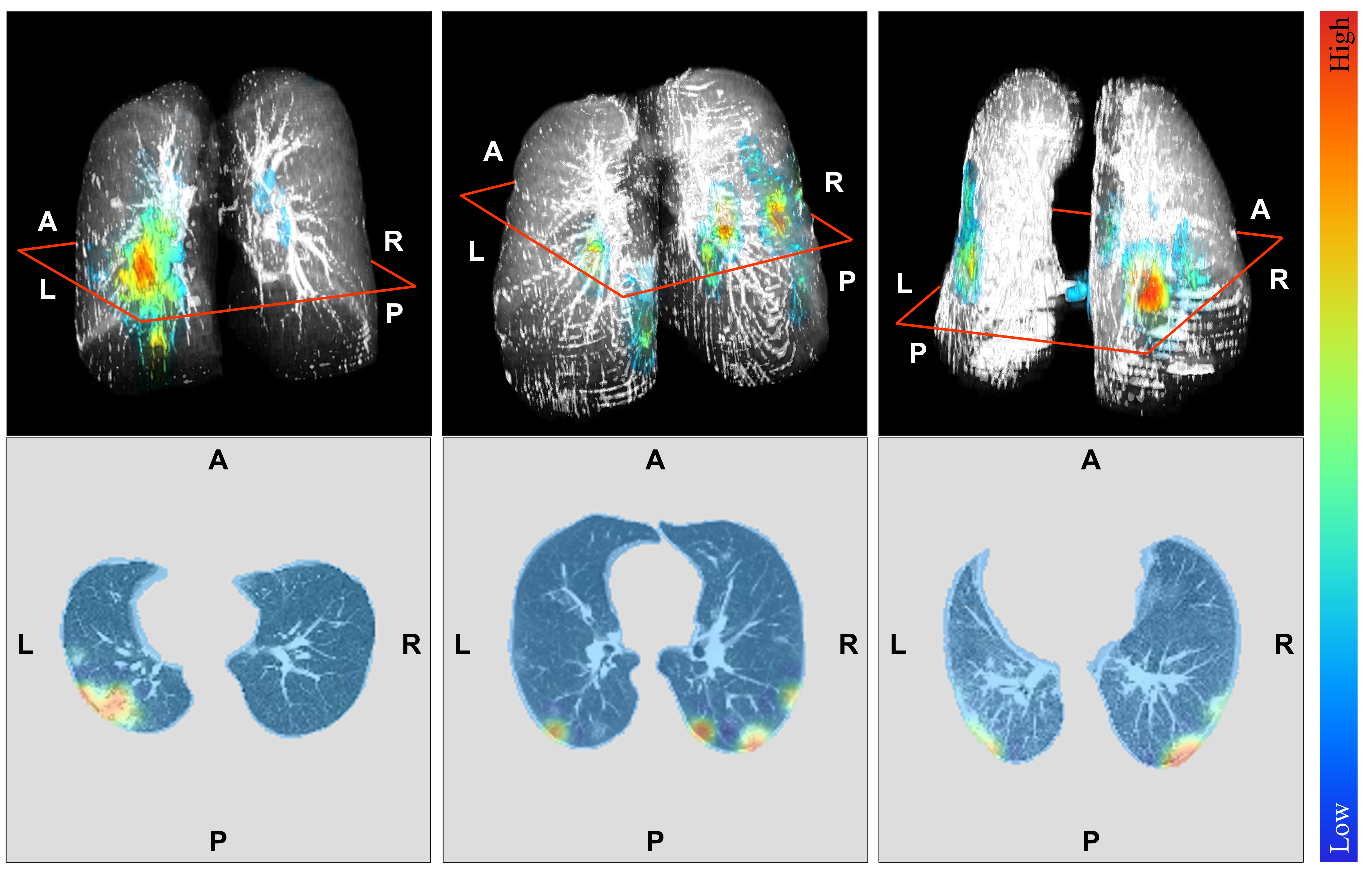}
    \caption{
      Examples of our suspicious-region identification in three testing data:
      in-house (left), COVID-CT-MD (middle), and COVID-CTset (right).
      Upper images show the thresholded $\bm{S}$ of
      Eq. (\ref{eq:PosiGradCAM3D})  superimposed on the CT volume.
      Lower images are corresponding axial slices before thresholding.
      We selected the slices capturing the center of the most significant lesion in each case.
    }\label{fig:Visualization2}
\end{figure}

From the in-house test data, we picked 47 typical cases whose radiology reports
contained information on the suspected COVID-19 lesions and
their lobe-level locations for evaluation.
For the selected cases, we evaluated a coincidence of actual lesions and
CNN-identified suspicious regions.
We checked the lesion location based on the radiology report
and intensities of the heatmap there.
If the intensities were higher than 0.1 (excluded level),
we regarded the lesion as correctly identified.
The evaluation is based on lobe- and case-level.
By using the total number of lesions $N_\mathrm{total}$ and
the number of identified lesions $N_\mathrm{id}$,
we computed identification rates (IR) for lobe- and case-levels as
$\mathrm{IR} = N_{\mathrm{id}}/N_\mathrm{total}$.
Figure \ref{fig:Visualization2} is an example of the suspicious-region identification on the three datasets.
Table \ref{tab:IdentificationRate} shows the accuracy of the suspicious-region identification. 
\begin{table}[tb]
    \caption{Comparative evaluation of COVID-19 classification.
      We compare AUCs of COVID-19 classification among the six methods.}
    \label{tab:AUCs}
    \centering
    \begin{tabular}{l||r|r|r}
      \hline
      & In-house & COVID-CT-MD & COVID-CTset \\
      \hline
      3D CNN & 0.838 & 0.853 & 0.779  \\
      3D CNN + attn. block & 0.906 & 0.932 & 0.900 \\
      2.5D CNN & 0.881 & 0.930 & 0.855 \\
      2.5D CNN + OEN & 0.869 & 0.929 & 0.889 \\
      2.5D CNN + 5 attn. blocks & 0.732 & 0.725 & 0.685 \\
      \bf{Proposed} & \bf{0.910} & \bf{0.946} & \bf{0.916} \\
      \hline
    \end{tabular}
\end{table}

\begin{table}[tb]
    \caption{Evaluation of suspicious-region identification.
    We evaluated the identification accuracy of five regions:
    left upper lobe (LUL), left lower lobe (LLL), right upper lobe (RUL),
    right middle lobe (RML), and right lower lobe (RLL).}
    \label{tab:IdentificationRate}
    \centering
    \begin{tabular}{l||rrrrr|r}
      \hline
      & \multicolumn{1}{|c}{LUL} & \multicolumn{1}{c}{LLL} &
      \multicolumn{1}{c}{RUL} & \multicolumn{1}{c}{RML} &
      \multicolumn{1}{c|}{RLL} & \multicolumn{1}{c}{Case-level}\\
      \hline
      Number of total & 21 & 33 & 23 & 11 & 29 & 47\\
      Number of identified & 2 & 22 & 9 & 2 & 27 & 39\\
      \hline
      Identification rate & 9.52\% & 66.7\% & 39.1\% & 18.2\% & 93.1\% & 83.0\%\\
      \hline
    \end{tabular}
\end{table}

\section{Discussions}
As shown in Table \ref{tab:AUCs} and Fig. \ref{fig:ROCCurves}, our proposed CNN
achieved the best performances among six methods on three datasets.
The results suggest that the 2.5D architecture with the 3D attention
mechanisms is effective to improve the model's generalization ability. 
Especially, the improvement in the specificity is evident. 
This implies that attention-guided features express key patterns of typical COVID-19 well.
On the other hand, the model with multiple attention blocks decreased its
classification performance from the model without attention blocks.
In this model, the attention mechanisms might perform over selections of
features in shallow layers and fail in feature extraction of key 3D patterns. 

The visualization in Fig. \ref{fig:Visualization2} implies that 
our method can identify ground-glass opacities (GGOs) and consolidations.
The method tended to focus on lesions located in posterior regions and lower lobes.
These findings and regions match with previously reported common findings and
sites of COVID-19 \cite{COVIDFindings2,COVIDFindings3}.
In Table \ref{tab:IdentificationRate},
the proposed method achieved the case-level IR over 80\%,
even though the IR of the left lower lobe is lower than the right lower lobe.
We can also interpret the low IRs for the upper and middle lobes
as a result of the classification mainly focusing on lesions in the lower lobes.
However, typical COVID-19 cases often have bilateral lesions \cite{COVIDFindings2,COVIDFindings3}.
Since our method was able to identify lesions in at least either lung,
these results suggest that our attention- and positive-gradient-based
visualization is acceptable for identifying suspicious regions in a practical application scenario.

As the current limitation, identifying weak and highly-localized GGOs is still challenging for the proposed method.
We will refine this point by adopting a larger input size of our architecture to capture more tiny key patterns. 

\section{Conclusions}
We proposed a suspicious region identification method for the COVID-19 cases
on chest CT volumes by an abnormality-sensitive activation mapping.
As a result of the experiments,
our method showed stable classification performances on three datasets.
The method identified GGOs and consolidations in posterior regions and lower lobes as suspicious regions.
Since these abnormalities are in agreement with the clinically important findings,
we conclude that the proposed method is acceptable for the diagnostic assistance.

\subsubsection{Acknowledgements}
Parts of this research were supported by 
the NICT, Grant Number 222A03, the JST CREST, Grant Number JPMJCR20D5,
the JSPS KAKENHI, Grant Number 26108006 and the JST SPRING, Grant Number JPMJSP2125.
The author (RT) would like to take this opportunity to thank the ``Interdisciplinary Frontier
Next-Generation Researcher Program of the Tokai Higher Education and Research System.''


\end{document}